# Largest Particle Simulations Downgrade the Runaway Electron Risk for ITER

Jian Liu[1*], Hong Qin[1,2*], Yulei Wang[1*], Guangwen Yang[3,4], Jiangshan Zheng[1], Yicun Yao[1], Yifeng Zheng[1], Zhao Liu[4], Xin Liu[5]

1. School of Nuclear Science and Technology and Department of Modern Physics, University of Science and Technology of China, Hefei, Anhui 230026, China
2. Plasma Physics Laboratory, Princeton University, Princeton, New Jersey 08543, USA
3. Department of Computer Science and Technology, Tsinghua University, Beijing 100084, China
4. National Supercomputing Center in Wuxi, Wuxi 214072, China
5. National Research Center of Parallel Computer Engineering and Technology
* These authors contributed equally to this work.

**Fusion energy will be the ultimate clean energy source for mankind. The ITER device under construction by international partners will bring this goal one-step closer [1,2]. Before constructing the first prototype fusion power plant, one of the most visible concerns that needs to be addressed is the threat of deleterious runaway electrons (REs) produced during unexpected disruptions of the fusion plasma [3-7]. Massive REs can carry up to 70% of the initial plasma current in ITER, and understanding their dynamical behavior is crucial to assess the safety of ITER. However, the complex dynamics of REs in a realistic fusion reactor is almost impossible to simulate numerically because it requires efficient long-term algorithms and super-large scale computing power [8,9]. In the present study, we deploy the world's fastest supercomputer, Sunway TaihuLight [10,11], and the newly developed relativistic volume-preserving algorithm [12] to carry out long-term particle simulations of $10^7$ sampled REs in 6D phase space. The size of these simulations is in the range of $10^{18}$ particle-steps, the largest ever achieved in fusion research. Previous studies suggest that REs can be accelerated to 350MeV or higher in ITER, and randomly strike the first wall of the reactor to cause grave damage [13-15]. Our simulations show that in a realistic fusion reactor, the concern of REs is not as serious as previously thought. Specifically, REs are confined much better than previously predicted and the maximum average energy is in the range of 150MeV, less than half of previous estimate. As a consequence, most of the energy carried by these electrons will be released through the benign process of synchrotron radiation without causing damage to the first wall. These simulations on Sunway TaihuLight ease the concern over REs, and give scientists more confidence in the outcome of ITER and Chinese Fusion Engineering Test Reactor (CFETR), which is the post-ITER device currently being designed.**

As the ITER project is making great progresses, there are still some design and operation challenges on the way to harvest fusion energy. Massive deleterious REs produced during disruption and fast shutdown become the most visible concern. It is believed that large amount of REs will be generated through avalanche multiplication in disruptions and accelerated to a high energy, and randomly strike the first wall of the reactor to cause grave damage. In ITER, REs can carry up to 70% of the initial plasma current under the large loop voltage of 1000V or higher. Under the influence of synchrotron radiation, poloidal flux variation, and magnetic field ripples, the maximum energy of each runaway electron (RE) is estimated to be 350MeV or higher in ITER. Though many important theoretical results have been achieved [16-20], the overall dynamic behavior of REs in a realistic tokamak configuration is still rather vague. To evaluate the impact of the REs on the safety of the device, a comprehensive study of statistical properties of RE dynamics is necessary.

Because REs are highly relativistic with negligible collisions, accurate descriptions for their dynamics should be based on kinetic models. Fluid descriptions, such as the Magneto-hydrodynamics model, cannot capture many important features of the REs. Unfortunately, due to extreme multi-scale and nonlinear nature of the runaway dynamics, it is difficult to accurately trace the trajectory of even one single RE in phase space for its full lifecycle, from its creation to acceleration, and to its landing on the first wall or slowing down. With highly relativistic velocities, the runaway dynamics involves many coupled characteristic timescales spanning 11 orders of magnitude. Gyro-center approximation is no longer valid for REs with high energy [8]. Therefore, long-term simulation up to $10^{11}$ time steps is required to reveal the multiscale dynamics in the complex geometry. This poses a challenge for qualified numerical methods. Traditional algorithms, such as the Runge-Kutta method, cannot be applied to such long-term simulations. This is because for these algorithms, the truncation errors at different time steps accumulate coherently, and long-term numerical solutions are dominated by large numerical error, and thus not trustworthy. Efficient algorithms suitable for massively parallel supercomputers with long-term accuracy and fidelity need to be developed. On the other hand, a large number of sample points of REs in 6D phase space are required. Because the gyro-symmetry is broken with the failure of the gyro-center approximation and the toroidal symmetry is broken due to the realistic reactor configuration with the ripple field, at least ten values need to be sampled in each dimension to provide a satisfactory resolution in phase space, which means that at least $10^6$ to $10^7$ REs need to be sampled. The total amount of computation for one simulation study requires up to $10^{18}$ particle-steps, corresponding to more than $10^{21}$ floating-point operations, which is much larger than the largest particle simulation ever achieved in fusion research. Meanwhile, Petabyte-level high-performance storage system with large-scale parallel I/O is also necessary for handling the simulation data. These requirements on both hardware and software make a comprehensive study of REs using particle simulations almost impractical.

Recently, breakthroughs in both hardware and software enabled the first ever panoramic particle simulation of REs in a realistic fusion reactor. On the hardware side, the Sunway TaihuLight supercomputer, world's first supercomputer reaching the 100PFlops landmark, came online early 2016 at the National Supercomputing Center in Wuxi (NSCC-Wuxi) [11]. The peak performance of the Sunway TaihuLight is 125 PFlops with a Linpack rating of 93 PFlops. It is currently world's fastest supercomputer on the latest Top500 list [10]. One major technological innovation of the Sunway TaihuLight supercomputer is its SW26010 many-core processor, which includes 4 management processing elements (MPEs) and 256 computing processing elements (CPEs). On the software side, a large-scale particle simulation code, Accurate Particle Tracer (APT), has been developed based on a series of structure-preserving geometric algorithms, such as the explicit volume-preserving algorithms and the explicit symplectic algorithms, for relativistic particles [21-25]. These innovative algorithms preserve the geometric structure and energy-momentum of the relativistic dynamics and guarantee the long-term accuracy of the simulations beyond $10^{12}$ steps in time.

In the present work, long-term simulations on 10 million cores of Sunway TaihuLight supercomputer with the relativistic volume-preserving algorithm have been carried out using the APT code. A resolution of $10^7$ in 6D phase space is used, and simulations are performed for $10^{11}$ steps. Statistical properties of the runaway dynamics in both the realistic ITER configuration and the ideal configuration without considering magnetic field ripples are studied for comparison.

The runaway dynamics in the ITER device is illustrated in Fig. 1a, and the evolution of density distribution of the REs in the poloidal plane is displayed in Figs. 1b-1g. Figs. 1b, 1c, and 1d shows that the REs in the ideal configuration keep drifting outwards and finally strike the first wall of the device. This is consistent with previous theories and can be explained as the effect of canonical angular momentum conservation [26,27]. The loss rate of the REs is 2.6% at t=0.4s. However in the realistic configuration, the runaway beam is confined very well to the core region, and no RE is found to impact the wall, see Figs. 1e, 1f, and 1g. The runaway transit orbits shrink into the core because electrons experience an average focusing force caused by the toroidal field ripples as they travel in the toroidal direction for many turns. In this context, the toroidal ripple field behaves similarly as a periodic focusing lattice in charged particle accelerators and rings [28]. The enhanced confinement of REs greatly reduces the risk of direct impact of the runaway beam on plasma facing components.

The average energy of the REs is plotted in Fig. 2a for both the realistic and the ideal configurations. For the ideal case (red curve), the average energy increases to the synchrotron energy limit at 320MeV. For the realistic case (blue curve), the maximum energy is reduced to the level of 150MeV due to two effects produced by the ripple field, the increased collisionless pitch-angle scattering [8] and the harmonic resonances [13]. The increased pitch-angle scattering results in an enhanced synchrotron radiation, as demonstrated in Fig. 2b where the ratio of total runaway radiation power in the realistic configuration to that in the ideal configuration is plotted. After 0.1s, there is a rapid increase of the ratio, which corresponds to the separation point of the two curves in Fig. 2a. After 0.3s, the ratio falls back to 1 gradually, since the average energy in both cases reach steady state and all the work by loop electric field is dissipated by radiation. The enhancement of radiation is still in effect, i.e., the 150MeV electrons in the realistic configuration radiate as much as the 320MeV electrons in the ideal configuration. The enhanced pitch-angle scattering by the ripple field is evident from the evolution of pitch-angle distribution plotted in Fig. 3. In the ideal configuration, the pitch-angle converges to an equilibrium value set by the collisionless neoclassical scattering, see Fig. 3a. The ripple field in the realistic configuration induces a much larger collisionless pitch-angle scattering effect, which is responsible for the significantly increased average and spread of the pitch-angle distribution shown in Fig. 3b.

The damping effect on runaway energy due to resonances with the ripple field was first discovered by Laurent and Rax [13], who predicted that the nth toroidal harmonic of the ripples can put a damp on runaway energy through the resonance at $E_n=BeRc/nN$. For ITER parameters, the second harmonic resonant energy is $E_2$=274MeV and the third is $E_3$=183MeV. However, for a distribution of REs in 6D phase space, it is difficult to estimate analytically the average runaway energy under these resonances even in a simplified model. In addition, because the gyro-center approximation is not valid and the collisionless pitch-angle scattering is significant, large-scale numerical simulation is the only feasible method to trace the complex multiscale dynamics of REs and calculate the energy distribution in the realistic configuration. The evolution of energy distribution of the REs is displayed in Fig. 4. For the case of the realistic configuration, the energy distribution exhibits complex bifurcation phenomena. The first obvious bifurcation takes place at 0.07s, and a wide spread of energy distribution occurs at 0.1s. There are several energy boundaries as well. A large portion of the REs are bounced down at the first boundary at 175MeV after 0.15s, and a small percentage penetrate the first boundary and are bounced back at the second boundary at 258MeV. It can be inferred that the two boundaries in Fig. 4b correspond to the $E_2$ and $E_3$ in Laurent and Rax's theory, though the values are somewhat different. Furthermore, there is also a weak boundary at 130MeV, which is responsible for the distribution spread at 0.1s. These colorful details revealed by the big data generated by the large-scale simulations are not accessible by previous theoretical or numerical models. In the present study, the impact of the ripple field mainly comes from its n=3 harmonic component, even though the third harmonic ripple field is only 0.06% of the toroidally symmetric field at the edge of plasma. For the ideal configuration, the evolution of the distribution is rather simple. The width of Gaussian distribution is always small, which shows good monochromaticity of the runaway beam. The peak of distribution grows monotonically, and the evolution of energy distribution in the ideal configuration is almost the same as the evolution of its average energy, see the red curve in Fig. 2a.

In addition to the significant reduction in the average energy, the surprisingly large spread in REs' energy observed in the simulation for the realistic configuration also reduces the threat of REs. It provides an effective stabilizing effect for possible instabilities of the runaway beam through the well-known process of Landau damping.

For a comparison, we note that in charged particle accelerators, a fraction of 1% spread in beam momentum is able to stabilize many detrimental beam instabilities, such as the electron-ion two-stream stabilities [29].

In conclusion, large-scale long-term simulations on world's fastest computer reveal that REs in the realistic field configuration of ITER pose a much smaller threat to the device than previously predicted by overly simplified theoretical and computational models. More importantly, it is discovered that a small percentage of the 3D ripple components of the magnetic field are able to significantly reduce the RE risk for ITER. The Sunway TaihuLight supercomputer and the APT simulation code can be effectively utilized to design optimized magnetic coils to control REs for ITER and post-ITER devices, such as the CFETR.

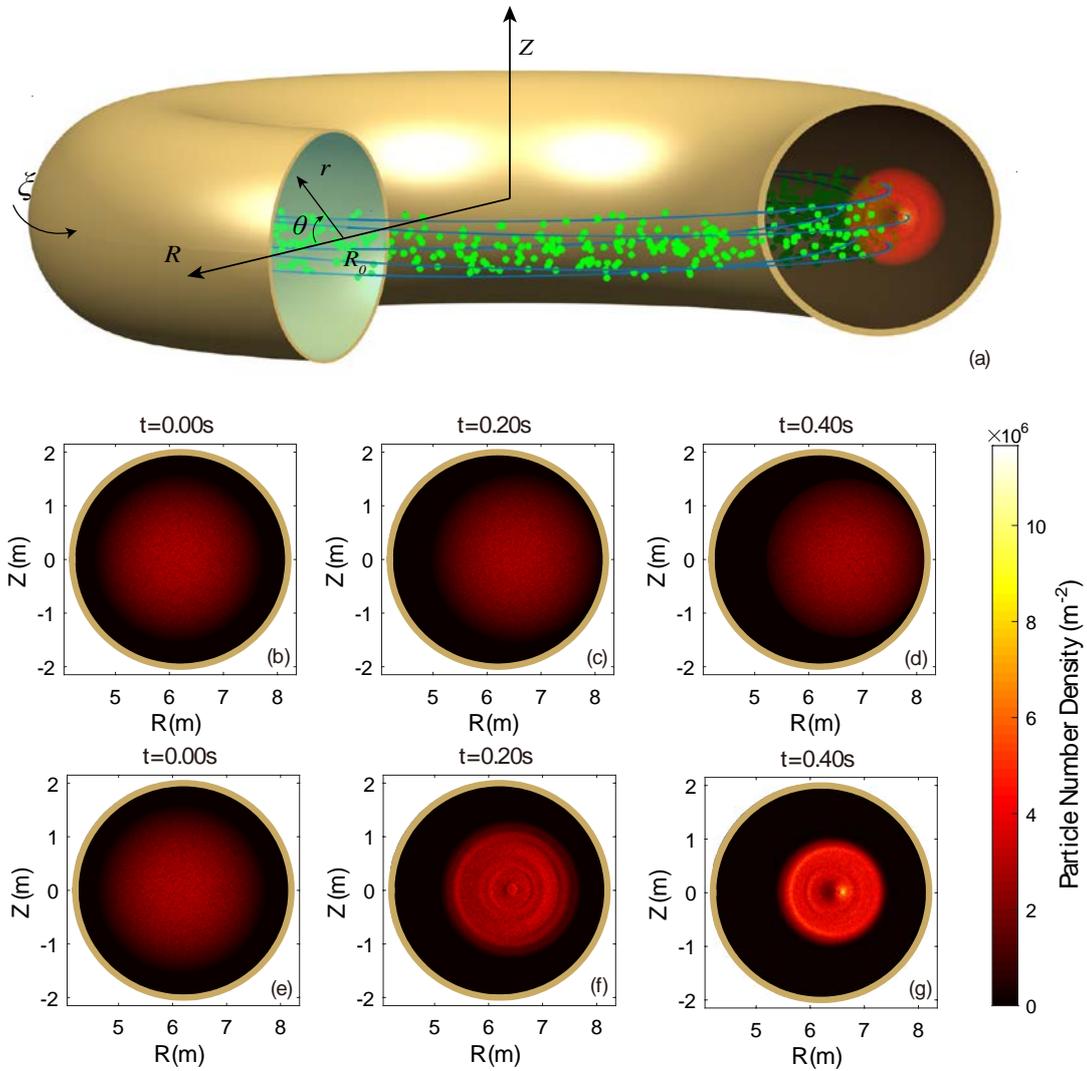

**Figure 1: (a) Illustration of runaway dynamics in a tokamak and evolution of the RE distribution in a poloidal cross-section in (b)-(d) the ideal configuration and (e)-(g) the realistic configuration.** The color bar indicates the number density of sampled REs within

the poloidal cross-section, which are calculated using a resolution of 8mm by 8mm. In the ideal configuration, the REs drift outwards to the first wall of the device with a slightly increased spread. In the realistic configuration the REs are concentrated to the core region due to the periodic focusing effect produced by the ripple field. A ring-like pattern appears at t=0.2s. At t=0.4s, there exists a dipole structure in the center of the beam. In the ideal configuration, 0.018% of the REs strike the first wall near the position z=0m, R=8.2m at t=0.2s, and at t=0.4s the loss rate increase to 2.6%. In the contrary, there's no loss of REs in the realistic configuration. The much improved confinement in the realistic configuration enables more runaway energy released through radiation instead of direct impact on the first wall.

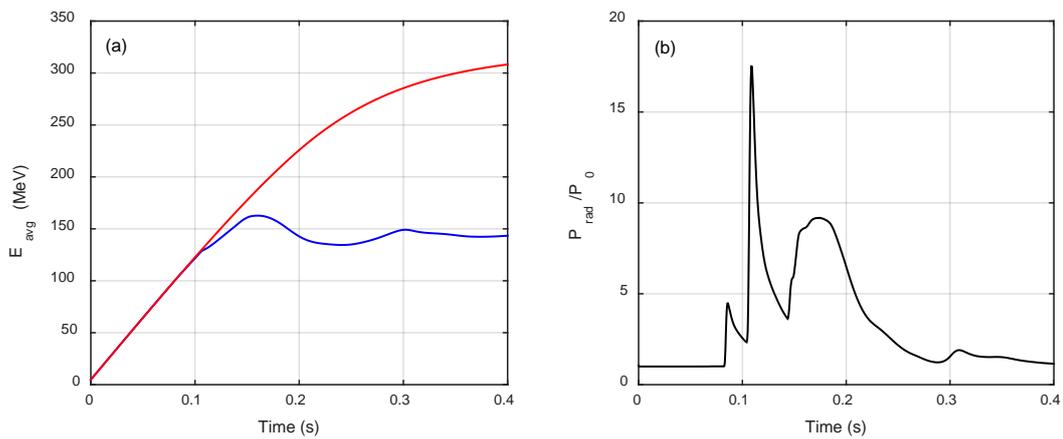

**Figure 2: (a) Evolution of average runaway energy in the realistic (blue) and ideal (red) configurations. (b) Evolution of the ratio of total runaway radiation power in the realistic configuration to that in the ideal configuration.** Subfigure (a) shows that the maximum runaway energy in the realistic configuration is reduced to half of that in the ideal configuration. Subfigure (b) shows the greatly enhanced radiation power induced by magnetic field ripples.

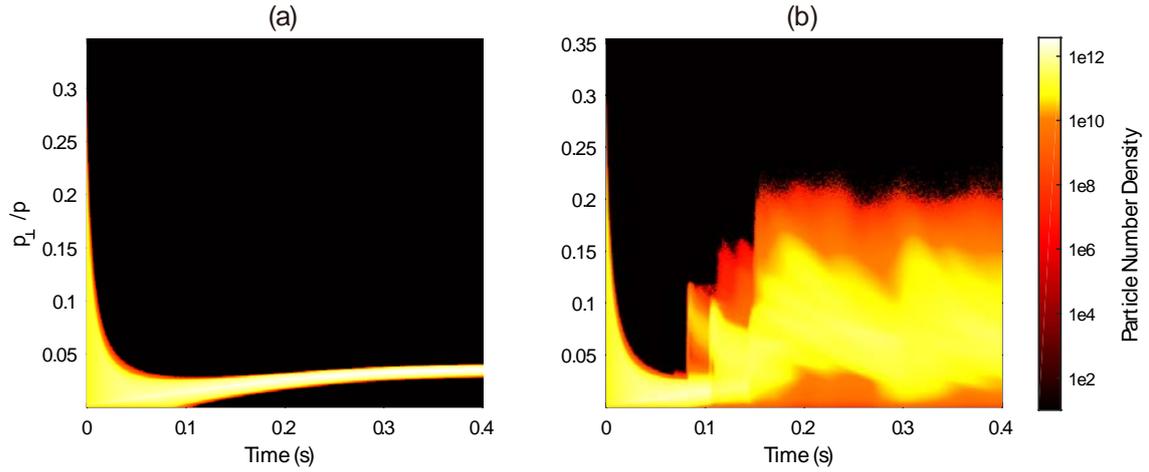

**Figure 3: Evolution of the pitch-angle distribution of the REs in (a) the ideal and (b) the realistic configurations.** The collisionless pitch-angle scattering due to the ripple field in the realistic configuration generates a much larger average and spread for the pitch-angle distribution.

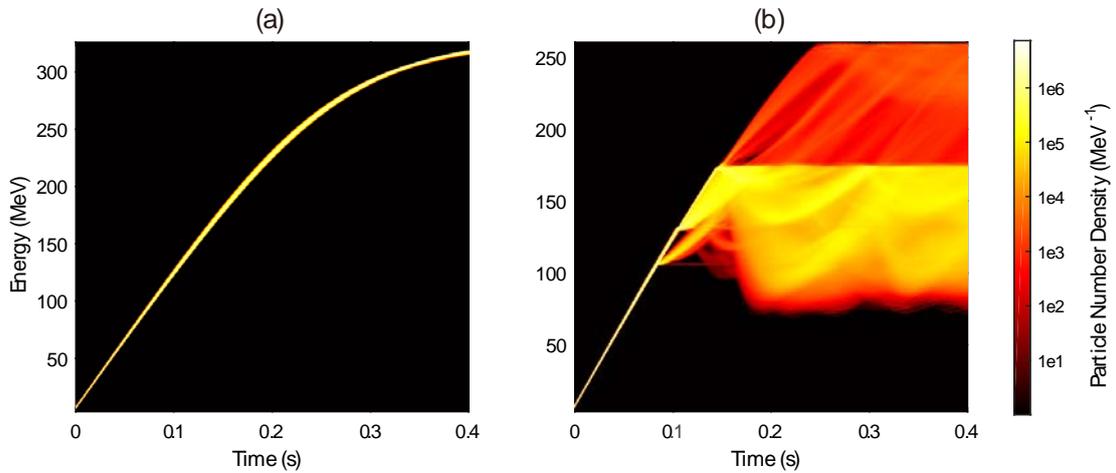

**Figure 4: Evolution of the energy distribution of the REs in (a) the ideal and (b) the realistic configuration.** In the ideal configuration, the energy distribution grows monotonically with time with a very small spread. Its trend is similar to the energy evolution of a single runaway electron in the ideal case. The energy distribution in the realistic configuration exhibits complex behavior, characterized by a much reduced average value and a large spread.

**Author Contributions**

J. L. and H. Q. conceived the large-scale particle simulations for runaway dynamics, the design of physical models, and the advanced geometric numerical methods. Y. W. implemented the APT code on the Sunway TaihuLight supercomputer and carried out the simulations. Y. W., J. Z, Y. Y., and Y. Z. performed the data analysis and visualization. G. Y., Z. L., and X. L. provided hardware support, parallel optimization, and the optimization of the compiler library for CPEs.

**ACKNOWLEDGMENTS**

This research was supported by National Magnetic Confinement Fusion Energy Research Project (2015GB111003, 2014GB124005), the National Natural Science Foundation of China (NSFC-11575185, 11575186, and 11305171), JSPS-NRF-NSFC A3 Foresight Program (NSFC-11261140328), and the GeoAlgorithmic Plasma Simulator (GAPS) Project.

## METHODS

**Simulation Model.** The dynamics of each sampled runaway electron in 6D phase space (**x**, **p**) is governed by Newton's equation,

$$\frac{d\mathbf{x}}{dt} = \mathbf{v},$$

$$\frac{d\mathbf{p}}{dt} = -e(\mathbf{E} + \mathbf{v} \times \mathbf{B}) + \mathbf{F_R},$$

$$\mathbf{p} = \gamma m_0 \mathbf{v},$$

where the external magnetic field **B** consists of toroidally symmetric field $\mathbf{B_0}$ and the ripple field $\delta\mathbf{B}$, **E** is the loop electric field, and

$$\mathbf{B_0} = -\frac{B_0 R_0}{R}\mathbf{e}_\xi - \frac{B_0\sqrt{(R-R_0)^2 + z^2}}{qR}\mathbf{e}_\theta,$$

$$\delta\mathbf{B} = \delta B \mathbf{e}_r,$$

$$\mathbf{E} = E_l \frac{R_0}{R}\mathbf{e}_\xi.$$

Here, we use the cylindrical coordinate system $(R, \xi, z)$. The vectors $\mathbf{e}_r$, $\mathbf{e}_\theta$ and $\mathbf{e}_\xi$ are respectively unit vectors in the radial, poloidal and toroidal direction, $R_0$ is the major radius, $q$ denotes safety factor, $E_l$ is the strength of loop electric field, and $B_0$ is the magnitude of background magnetic field. The effective electromagnetic radiation drag force $\mathbf{F_R}$ is

$$\mathbf{F_R} = -\frac{P_R}{v^2}\mathbf{v},$$

where $P_R$ is the radiation power determined by

$$P_R = \frac{e^2}{6\pi\epsilon_0 c}\gamma^6\left[\left(\frac{\mathbf{a}}{c}\right)^2 - \left(\frac{\mathbf{c}}{c} \times \frac{\mathbf{a}}{c}\right)^2\right].$$

Here, $\epsilon_0$ is the permittivity of vacuum, c is the speed of light in vacuum, and $\mathbf{a} = d\mathbf{v}/dt$ is the acceleration.

Keeping only several leading harmonics of magnetic field ripples, m=0,1 and n=1,2,3, we have

$$\delta B(r, \theta, \varphi) = \sum_{m=0, n=1}^{m=1, n=3} \delta B_{mn}(r) \cos(m\theta) \cos(nN\varphi),$$

$$\delta B_{0n}(r) = \delta B_{1n}(r) = \eta_n B_0 \frac{r^2}{a^2}.$$

Here, $r$, $\theta$, and $\varphi$ are the radial, poloidal and toroidal coordinates, N is the number of toroidal coils, and $\eta_n$ denotes the relative strength of n-th order ripple in the outboard plasma region. The following set of parameters for ITER is used [2,30]: $B_0$=5T, $q$=2.5, $R_0$ =6.2m, $a$=2m, $N$=18, $E_l$=4V/m, $\eta_1$=1%, $\eta_2$=0.2%, and $\eta_3$=0.06%. The initial distribution of REs on the poloidal plane are sampled according to the quadratic

profile

$$f(R,z) = f_0 \left[1 - \frac{(R-R_0)^2 + z^2}{r_b^2}\right],$$

where $r_b$ is the outer boundary of the initial distribution. In the present study, we take $r_b$=1.6m. The initial toroidal angle $\varphi_0$ and poloidal angle $\theta_0$ are sampled uniformly from 0 to $2\pi$. The initial kinetic energy is sampled as a normal distribution with expectation μ=4.75MeV and standard deviation σ=0.25MeV, the initial pitch-angle is uniformly sampled from 0 to 0.3, and the initial gyro-phase is distributed uniformly from 0 to $2\pi$.

**Geometric Algorithm.** Because the multiscale nature of runaway electron dynamics, more than $10^{11}$ time steps of simulation are required. This type of long-term simulation cannot be carried out using traditional algorithms. The relativistic volume-preserving algorithm (RVPA) [21-25] is adopted in this large-scale long-term simulation research to guarantee the long-term numerical accuracy and stability. The RVPA with radiation force is constructed as

$$\mathbf{x}_{k+\frac{1}{2}} = \mathbf{x}_{k-\frac{1}{2}} + \Delta t \frac{\mathbf{p}_k}{\sqrt{m_0^2 + \mathbf{p}_k^2/c^2}},$$

$$\mathbf{p}_{k+1} = \mathbf{p}_k - e\Delta t \left( \frac{\mathbf{p}_k + \mathbf{p}_{k+1}}{\sqrt{4m_0^2 + \frac{\left(2\mathbf{p}_k + \Delta t \mathbf{E}_{k+\frac{1}{2}}\right)^2}{c^2}}} \times \mathbf{B}_{k+\frac{1}{2}} + \mathbf{E}_{k+\frac{1}{2}} \right) + \Delta t \mathbf{F}_{Rk},$$

where $\Delta t$ is the time step, $\mathbf{E}_{k+\frac{1}{2}} = \mathbf{E}\left(\mathbf{x}_{k+\frac{1}{2}}, t + \frac{\Delta t}{2}\right)$, $\mathbf{B}_{k+\frac{1}{2}} = \mathbf{B}\left(\mathbf{x}_{k+\frac{1}{2}}, t + \frac{\Delta t}{2}\right)$, and $\mathbf{F}_{Rk} = \mathbf{F}(\mathbf{a}_k, \mathbf{v}_k)$. The RVPA can be constructed using the splitting method through three steps: (1) split the original system into three incompressible subsystems, (2) find a VPA for each subsystem, and (3) combine the sub-algorithms into a desired VPA for the original system. The RVPA guarantees the numerical accuracy and fidelity of long-term simulations by preserving the corresponding geometric structure, i.e., the phase-space volume.

**APT and its implement on TaihuLight.** The particle simulations in this work are carried out using the Accurate Particle Tracer (APT) code, which is based on a variety of outstanding geometric algorithms for charged particle dynamics, such as symplectic algorithms and volume-preserving algorithms. The RVPA solver used in this study is one of APT's geometric solvers, which has been proved to possess long-term numerical stability and bound the global error. APT provides a powerful tool for the study of the multiscale runaway dynamics. In the present study, the C-language version of the APT code was successfully implemented on the Sunway

TaihuLight supercomputer.

     The Sunway TaihuLight supercomputer is equipped with more than 10 million cores, which provide a computing power over 100Pflops for the large-scale particle simulations of REs. In order for the APT code to fully utilize the computing resources of Sunway TaihuLight, three key optimizations of the APT code have been implemented, i.e., improving the communication efficiency between management processing elements (MPEs) and computing processing elements (CPEs), increasing the performance of mathematical compiler library for CPEs, and optimizing the efficiency of I/O operations to the storage system. Each SW26010 many-core processor has 4 MPEs together with 256 CPEs. Though MPEs and CPEs have shared RAM, the efficiency for a CPE to visit the shared RAM is low. To improve the performance, the local dynamic memories of CPEs are engaged. At the same time, the APT code is optimized to reduce the communications between MPEs and CPEs. All mathematical compiler libraries for CPE architecture have been specially optimized to enhance the computing performance of mathematical functions. In addition, the I/O efficiency has also been significantly improved for the output data up to PBs when simulating $10^7$ particles for $10^{11}$ steps. The local storage system of Sunway TaihuLight is used to improve the I/O speed and guarantee the stability for frequent large-scale parallel reading and writing operations.

# Extended Data

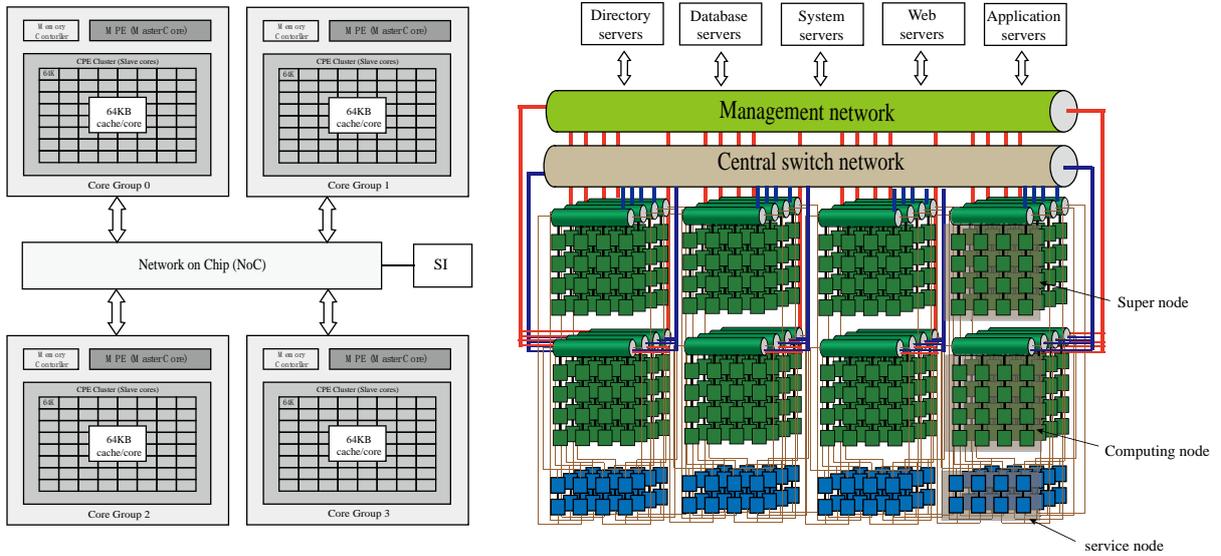

**Figure 5: General architecture of the Sunway Processor and the Sunway TaihuLight system.** The Sunway TaihuLight supercomputer is built with the innovative SW26010 many-core processors, each of which includes 4 management processing elements (MPEs) and 256 computing processing elements (CPEs). It is currently world's fastest computer on the latest Top500 list with a peak performance of 125 PFlops.

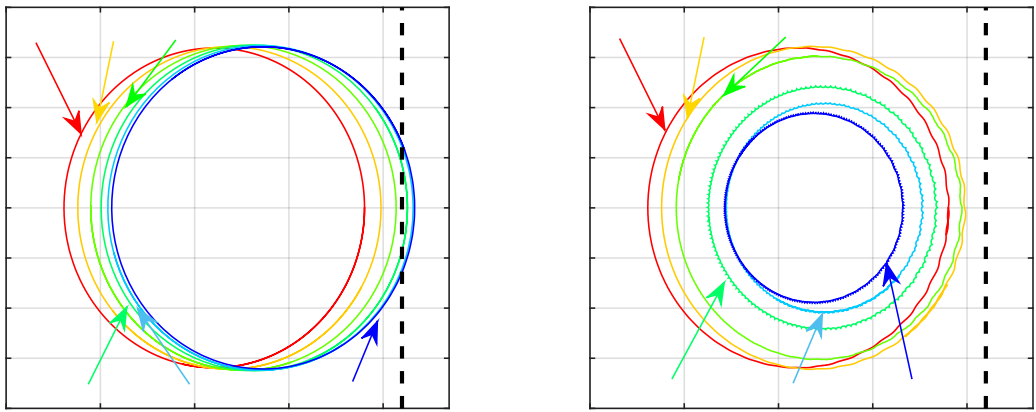

**Figure 6: Snapshots of the orbits for a sampled RE in the poloidal plane at different moments (a) in the ideal configuration and (b) the realistic configuration.** The initial condition of the sampled runaway electron is $r_0$=1.6 m, $\theta_0=\phi_0=0$, $p_{\parallel 0}$=7$m_0$c, and $p_{\perp 0}$=1$m_0$c. In the ideal configuration, because of the toroidal acceleration by the loop electric field, the runaway transit orbit drifts outwards. This neoclassical drift leads to the loss of the REs to the

first wall, which is indicated by the vertical dashed line. In the realistic configuration, the magnetic ripples provide a focusing effect on the RE, and it orbit shrinks to the core, resulting in the inward pinch of the density distribution shown in Fig. 1.

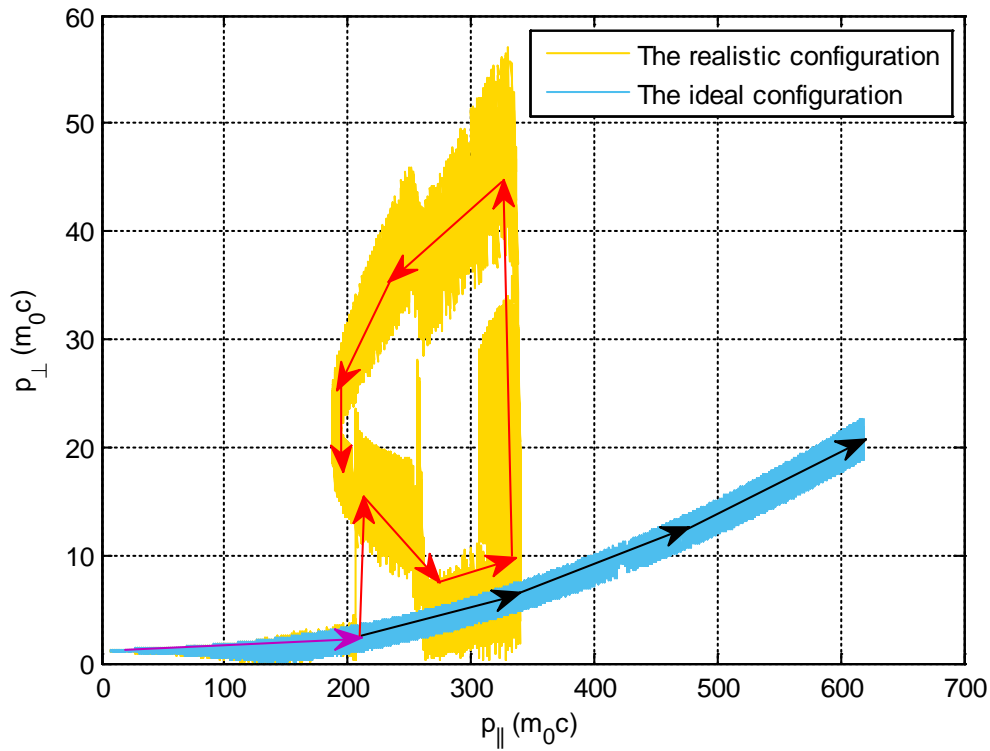

**Figure 7: Evolution of a sampled runaway electron in the momentum space in the ideal configuration (blue) and the realistic configuration (yellow).** Chronological order is indicated by arrows. At the beginning, the two curves overlap, showing similar behaviors. In the realistic configuration, the parallel momentum stops at about 170MeV and the perpendicular momentum starts to grow. Accompanied by strong oscillations, the momentum trajectory forms a cyclic pattern. The increase of the perpendicular moment enhances the synchrotron radiation and reduces the maximum energy of REs.